# Charging While Driving Lanes: A Boon to Electric Vehicle Owners or a Disruption to Traffic Flow

Shayan Bafandkar[a], Alireza Talebpour[a,*]

[a]*The Grainger College of Engineering, Department of Civil and Environmental Engineering, University of Illinois at Urbana-Champaign, Urbana, IL, USA, 61801*

## Abstract

Large-scale adoption of commercial and personal Electric Vehicles (EVs) is expected to significantly affect traffic flow dynamics, emissions, and energy consumption in the transportation sector. Range anxiety and challenges associated with charging EVs are among the key issues that reduce the adoption rate of EVs and, in turn, limit their system-level impacts. A promising solution to address these challenges is the introduction of charging while driving (CWD) lanes. Although technological advancements have made it possible to charge vehicles wirelessly while driving, introducing such lanes to the traffic stream can potentially disturb traffic flow and result in new congestion patterns. This study puts forward a framework to investigate the effects of CWD lanes on traffic flow, considering different market penetration rates (MPRs) of personal and commercial EVs. Different policies have been investigated to suggest the best design for CWD lanes. Results indicate that introducing CWD lanes can decrease overall traffic throughput and increase congestion due to additional lane-changing maneuvers by electric vehicles aiming to utilize the CWD lane. Although higher MPRs of EVs help stabilize traffic flow and reduce the number of shockwaves, speed disruption tends to increase in the CWD lane and propagate to adjacent lanes. Emission analyses show significant reductions (up to 63%) in pollution levels with increasing MPRs of personal and commercial EVs. Our analysis shows that while CWD lanes can facilitate the adoption of EVs, they can deteriorate traffic efficiency, emphasizing the importance of careful design and policy considerations.

---

*Corresponding author: `ataleb@illinois.edu`

## 1. Introduction

The electric vehicle (EV) market penetration rate (MPR) is increasing, and their market share is predicted to reach 50% by 2030 [1]. Despite these trends, the long-predicted electrification revolution in transportation is still facing certain challenges. Range anxiety and limited availability of charging stations are among the key factors prohibiting the large-scale adoption of EVs (both in personal and commercial segments). One potential solution to deal with these challenges is introducing charging while driving (CWD) lanes. CWD lanes can convert roads to charging infrastructure [2]. While wireless electrification technology can be traced back to the $19^{th}$ century, when Nikola Tesla developed a wireless light bulb [3], CWD lanes have only recently emerged and been implemented in several cases, mainly for testing and demonstration purposes. A few examples of such deployments are a 15-mile charging lane in Gumi, South Korea, for bus in-motion charging [4], a 2 km test track in Berlin constructed by Scania and Siemens [5, 6], a 2-mile charging system developed by Siemens and Volvo on a highway stretching between Los Angeles and Long Beach [7], a wireless charging track in Gothenburg, Sweden [8], and a charging lane trial that began in 2015 by Highways England [9, 10].

Various aspects of the deployment of CWD infrastructure have been explored in the literature. Chen et al. [11] explored the competitiveness of CWD lanes and showed that these lanes are economically feasible even in their current forms and are even more profitable than charging stations in the private sector. Accordingly, this technology could offer major benefits as it can remove bulky cable connections that might introduce unnecessary risk to the users, omit waiting times at charging stations, and last as long as normal pavement [12, 13]. However, along with all these benefits come some concerns that have not been extensively studied in the literature. Particularly, despite their potential impacts on traffic flow dynamics and operational efficiency, effective design procedures for CWD lane layouts remain under-investigated [14].

For instance, CWD lanes would cause EVs and electric drayage trucks (ETs) to perform additional lane-changing maneuvers to reach these lanes. Moreover, operational constraints would cause vehicles on CWD lanes to reduce their speeds for charging. These impacts could vary depending on the vehicle type and capabilities. For instance, automated vehicles (AVs) tend to handle speed variations and lane-changing maneuvers more efficiently [15], and they can benefit from CWD lanes since the majority of AVs available in today's market run on EV platforms [16]. On the other hand, human-driven vehicles (HVs) and diesel drayage trucks (DTs) may experience flow disruptions without gaining any benefit from CWD lanes. Therefore, it is important to consider AVs as well as other vehicle types in this study to understand the extent to which the benefits and drawbacks of CWD lanes apply.

Accordingly, the main contributions of this study are twofold: (1) It introduces a microscopic-level modeling approach to assess the potential impacts of CWD lanes on traffic flow dynamics. In order to fully assess the impacts of CWD lanes, this study considers the interaction between HVs, AVs, DTs, and ETs; and (2) It establishes a benchmark for different policies related to CWD lane placement and configuration on both flat and uphill roadways. This research increases our understanding of how different CWD placement and configuration policies change traffic patterns, emissions, and energy consumption.

The remainder of this paper is organized as follows: We present a background on CWD lanes in the next section. Then, the details of the simulation framework are described. A detailed simulation study of several scenarios in a mixed environment is presented next, including the evaluation of effects on traffic flow dynamics, congestion, throughput, and emissions. This section is followed by a comprehensive analysis of different policies to identify the proper implementation layout of CWD lanes. Finally, the paper concludes with summary remarks and future research needs.

## 2. Background

To analyze the effects of introducing CWD infrastructure on traffic flow, one has to investigate the reasons for confining the adoption of EVs in the first place, as CWD lanes are specifically designed to benefit EVs. For EVs to be a convincing alternative to traditional fuel-powered vehicles, there needs to be sufficient coverage of charging stations to serve cities in the same way that fuel stations serve traditional vehicles [17]. However, there is significant inequality between the demand for and supply of public charging stations (PCS), with many major cities lacking adequate charging infrastructure for daily EV use [18, 19]. For instance, in Shanghai, 80% of households share only 10% of PCS accessibility [20]. To further complicate matters, covering this gap requires a major investment in constructing charging facilities. Unfortunately, typical charging stations also come with their own challenges: users tend to develop a negative perception of the time spent at charging stations [21], they are prone to operational failures in the presence of disruptions [22], and many of the charging stations will hardly be profitable economically [23]. Moreover, different types of vehicles may need different kinds of chargers. For instance, the trucking industry has unique charging needs, requiring potentially dedicated infrastructure investments [24]. All the aforementioned challenges negatively affect users' intention to adopt EVs [25].

Given that our driving environment comprises various types of vehicles, including AVs and HVs, we should also cover the effect of introducing AVs into the flow of HVs. At the operational level, connectivity and automation technologies are intended to enable vehicles to make safe and reliable decisions about acceleration choices and execute lane-changing maneuvers [15]. Therefore, AVs have the potential to reduce driving errors [26] as they can move with higher precision, higher stability (local and string stability [15]), and lower headway than HVs. These abilities heavily influence different aspects of traffic flow, including, but not limited to throughput, safety, energy efficiency, and speed. In fact, it has been shown that at high penetration rates, AVs could reduce crashes, velocity differences, instability, and stop-and-go traffic conditions; hence, improving traffic flow safety, throughput, efficiency, and speed in a mixed traffic scenario [15, 27, 28, 29, 30, 31, 32, 33]. AVs, many of which are EVs, will also change infrastructure requirements, as they need charging stations to operate. One of the now available technologies is the CWD lanes, which are competitive compared with charging stations for attracting drivers [11] and AV operators.

CWD lanes have become the focus of some recent studies, such as [34, 35, 36, 37], which investigate the technical challenges of implementing this technology. Since this is outside the scope of our study, a discussion on these studies is not included. Some of the impacts of installing CWD lanes on traffic flow have also been studied. He et al. [38] studied the adverse effects of deploying them in the Nguyen–Dupuis and Sioux Falls networks and demonstrated that the deployment of CWD lanes may reduce road capacity, which in turn affects travelers' route choice and probably decreases network traffic efficiency (i.e., increases total network travel time). Furthermore, the presence of CWD lanes may reduce budgets for constructing charging stations along highways. This could restrict AVs to using only CWD lanes, effectively turning them into fixed reserved lanes for AVs, a scenario that significantly increases congestion and leads to breakdown formation [39, 40]. Tran et al. [41] also focused on the network-level impacts of CWD lanes and developed a mixed-integer linear program to optimally deploy these lanes considering route choice behaviors. Chen et al. [11] examined the route choice behavior of EV drivers with the deployment of charging lanes under a network equilibrium state. Another study that covered this topic from the perspective of transportation network design and strategic planning level is Colovic et al. [42]. They developed a bi-level multi-objective network electrification design (BM-NED) model for assessing the adoption of the eHighway system. Their model considers multiple goals, including minimizing the total travel cost, infrastructure, and environmental costs and maximizing the average traffic density of OC hybrid trucks on electrified arcs. Moreover, several studies, such as [43, 44], have provided insights into the CWD lane length requirements under different traffic flow and demand scenarios.

To the best of our knowledge, most of the works related to CWD lanes from the transportation sector perspective focus on either the optimal configuration and allocation of these lanes into the system or the impacts of these lanes on route choice behavior and traffic assignment over the network. Our study is the first to try to develop a microscopic simulation framework to examine the impacts of CWD lanes on the traffic flow dynamics at a segment level. This is particularly important since any significant changes in the traffic flow pattern can influence the shape of the fundamental diagram and change the equilibrium state at the network level.

## 3. Methodology

To test the effects of CWD lanes on traffic flow dynamics, this study extended the microscopic simulation tool developed by Talebpour et al. [45] and Talebpour and Mahmassani [15]. This framework enables accurate simulation of HVs and AVs as well as both vehicle-to-infrastructure (V2I) and vehicle-to-vehicle (V2V) communications. As part of this study, models of DT and ET are also incorporated in this simulation framework.

### 3.1. Modeling Framework

As autonomy increases situational awareness beyond human capabilities, AVs' behavior is modeled based on van Arem et al. [46] and Reece and Shafer [47]. In this modeling framework, an AV's acceleration is determined based on the difference between current and intended speed as well as the distance and speed difference between the leading and target vehicles. The framework also considers sensor range and capabilities to ensure maximum safety at all times.

$$a_{\text{ref}_d} = k_a \cdot a_p + k_\nu \cdot (V_p - \nu) + k_d \cdot (r - r_{\text{ref}}) \tag{1}$$

where $a_p$ is the target vehicle's acceleration rate, $v_p$ the target vehicle's speed, $r_{ref}$ is the reference clearance to the target vehicle, and $k_a$, $k_v$, and $k_d$ are model parameters.

Following the approach presented in Talebpour and Mahmassani [15], to model HVs', DTs', and ETs' behavior, we adopted a deterministic acceleration modeling framework, Intelligent Driver Model (IDM) [48, 49]. IDM is versatile and widely used for different classes of vehicles. This model is a car-following model that captures how vehicles adjust their acceleration or deceleration based on the distance to the vehicle ahead relative to the desired speed ratio.

$$a_{IDM}^n(s_n, v_n, \Delta v_n) = a_n \left[ 1 - \left(\frac{v_n}{v_0^n}\right)^{\delta_n} - \left(\frac{s^*(v_n, \Delta v_n)}{s_n}\right)^2 \right] \tag{2}$$

$$s^*(v_n, \Delta v_n) = s_n^0 + T_n v_n + \frac{v_n \Delta v_n}{2\sqrt{a_n b_n}} \tag{3}$$

where maximum acceleration ($a_n$), comfortable deceleration ($b_n$), safe time headway ($T_n$), standstill gap ($s_n^0$), and delta ($\delta_n$) are parameters to be calibrated. Finally, the lane-changing behavior of all of the vehicles follows the Minimizing Overall Braking Induced by Lane Changes (MOBIL) [50, 51, 52] model. It is important to note that while more sophisticated lane-changing models exist (e.g., game theory based model of Talebpour et al. [53]), for the purpose of this study, MOBIL offers a robust and tractable lane-changing modeling framework that can be extended to various vehicle types.

*3.2. Model Calibration*

This section presents the details of the model calibration process and outcome for different types of vehicles.

*3.2.1. Third Geneartion Simulation Dataset*

Third Generation Simulation (TGSIM) Data is an extensive dataset collected from human-AV interactions at various levels of autonomy [54, 55]. The interaction between humans and AVs across different scenarios, environments, and levels of automation was captured by three videography methods: (1) fixed location aerial videography (a helicopter hovers over a segment of interest) over I–90/I–94 in Chicago, IL where data from SAE Level 2 AVs with active automated lane-changing functions mixed in the traffic flow were collected; (2) moving aerial videography (a helicopter follows the automated vehicles as they move through a much longer segment than in the first method) were used to collect four datasets on I–90/I–94 and I–294 in Chicago, IL, including SAE Level 1 and Level 2 AV fleets mixed in the traffic, and (3) infrastructure-based videography (multiple overlapping cameras located on overpasses create a comprehensive highway image) were used to collect one dataset near the George Washington University Campus in Washington D.C., from SAE Level 3 AVs and another dataset from SAE Level 2 AVs on I-395 in Washington D.C. [56]. Thus, this dataset offers an excellent opportunity to calibrate the parameters of HVs and DTs in a mixed traffic environment.

*3.2.2. Calibration Approach: Genetic Algorithm*

To utilize the above models effectively, we need to calibrate the models' parameters based on real-world observations. As our environment is comprised of both passenger cars and trucks, we calibrated IDM parameters for HVs and DTs based on the data from I-294 in Chicago, IL [54]. I-294 offers a large number of DTs compared to other locations in the TGSIM dataset. The genetic algorithm (GA) is used to calibrate the IDM parameters. The calibration process begins with 80 initial population sets, breeding through 40 generations with a mutation rate of 0.1 to find the set of parameters($\Theta_i$) that minimizes the following fitness function:

$$f(\Theta_i) = \frac{1}{1 + \sum_j \left( |p_{obs,j} - p_{sim,j}(\Theta_i)| + |v_{obs,j} - v_{sim,j}(\Theta_i)| \right)} \qquad (4)$$

where $p_{obs,j}$ and $v_{obs,j}$ are the positions and velocities observed at time $j$, and $p_{sim,j}(\Theta_i)$ and $v_{sim,j}(\Theta_i)$ are the simulated positions and velocities using the set $\Theta_i$. The sum of absolute differences between observed and simulated positions and velocities is minimized through this fitness function.

The outcome of calibration, except for $s_n^0$, is presented in Figure 1. This figure presents the distribution of $a_n$, $b_n$, and $T_n$ values for both HVs and DTs. The $s_n^0$ values are calculated by grouping the data into four different follower-leader interaction scenarios and then averaging the standstill gap for each grouping.

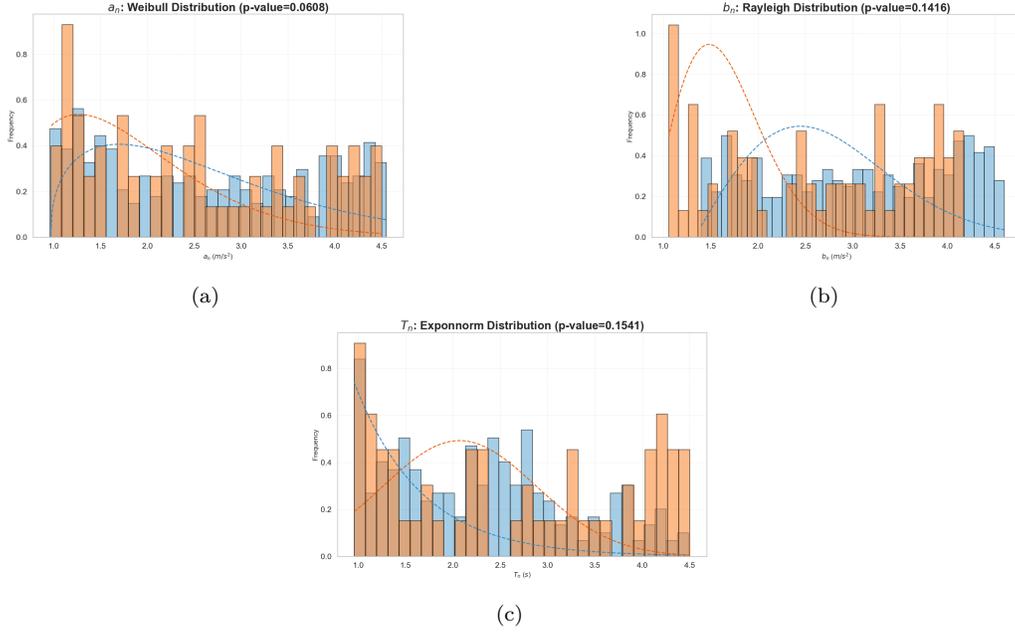

Figure 1: Distributions fitted for the IDM parameters using the TGSIM dataset [54]: (a) maximum acceleration ($a_n$), (b) comfortable deceleration ($b_n$), (c) safe time headway ($T_n$).

$$s_n^0 = \begin{cases} 2.66, & \text{if the follower is HV and the leader is also HV,} \\ 3.99, & \text{if the follower is DT and the leader is also DT,} \\ 2.73, & \text{if the follower is HV and the leader is DT,} \\ 2.74, & \text{if the follower is DT and the leader is HV.} \end{cases} \quad (5)$$

In Figure 1a, the fitted distribution for $a_n$ is shown. The blue bars represent HV data, accompanied by a dashed blue Weibull probability density function (PDF) with the following parameters: shape = 1.39, scale = 0.97, and location = 1.81. The orange bars represent DT data, paired with a dashed orange Weibull PDF that has a shape of 1.51, a scale of 0.59, and a location of 1.39. Figure 1b presents the fitted distribution for $b_n$. In this figure, the blue bars represent HV data with a dashed blue Rayleigh PDF, which has a scale parameter of 1.33. The orange bars denote DT data, accompanied by a dashed orange Rayleigh PDF with a scale of 0.84. Lastly, in Figure 1c, the fitted distribution for $T_n$ is depicted. The blue bars represent HV data with a dashed blue Exponnorm PDF characterized by a shape parameter of 2341.40, a location parameter of 0.50, and a scale parameter of 0.00. The orange bars illustrate DT data with a dashed orange Exponnorm PDF, which has a shape parameter of 0.22, a location parameter of 1.90, and a scale parameter of 0.79.

As ETs are mostly at the research and development stage, collecting real-world data from ET operations is challenging. Accordingly, we set the maximum acceleration of ETs to 1.3 m/s$^2$ based on the performance of the Tesla Semi, which can pull a fully loaded trailer and go from 0 to 100 km/hr in 5 s (making its acceleration rate 1.3 m/s$^2$) [57]. For the parameters of AVs and ETs (excluding acceleration rate), we adopted the calibration results from van Arem et al. [46] and Treiber [58], with the parameter values depicted in Tables 1 and 2, respectively.

Table 1: IDM Model Parameters for HVs, DTs, and ETs

| Parameter | HVs | DTs | ETs |
|---|---|---|---|
| $a_n$ | $a_{HV}$ | $a_{DT}$ | $1.3\,\frac{\text{m}}{\text{s}^2}$ |
| $b_n$ | $b_{HV}$ | $b_{DT}$ | $-2\,\frac{\text{m}}{\text{s}^2}$ |
| $T_n$ | $T_{HV}$ | $T_{DT}$ | $1.7\,\text{s}$ |
| $s_n^0$ | $s_{HV}^0$ | $s_{DT}^0$ | $2.0\,\text{m}$ |
| $\delta_n$ | 4 | 4 | 4 |

In the MOBIL lane-changing model [50, 51, 52], each vehicle's decisions are influenced by three main factors: incentive (whether a lane change improves the moving characteristics of a vehicle), politeness (whether a lane change negatively impacts the nearby traffic flow), and safety (whether the braking acceleration exceeds the safe deceleration rate). Based on the findings of Matcha et al. [59] for lane-changing behavior of vehicles in a mixed traffic environment, we used a safe deceleration rate of 4 m/s$^2$, politeness factor of 0.2, and a switching threshold of 0.1 m/s$^2$.

Table 2: Parameters for AVs in the Van Arem Model

| Parameter | Value |
|---|---|
| $a_{\max}$ | $4\,\text{m/s}^2$ |
| $b_{\max}$ | $-8\,\text{m/s}^2$ |
| $R_{\text{iso}}$[1] | $90\,\text{m}$ |
| $R_{\text{conn}}$[2] | $300\,\text{m}$ |
| $k$ | 0.3 |
| $k_a$ | 1 |
| $k_v$ | 0.58 |
| $k_d$ | 0.1 |
| $\tau_{\text{react}}$[3] | $0.5\,\text{s}$ |

*3.3. Vehicle Configurations and Assumptions*

Our simulation environment includes four vehicle categories: HVs, AVs, DTs, and ETs. Vehicles in each category are assumed to be uniform, with lengths of 5 m for HVs and AVs (as passenger cars) and 20 m for DTs and ETs. ETs were included in the fleet as they are expected to benefit significantly from CWD lanes, especially during long-haul trips. The vehicle generation process follows the method described in [45, 15, 60, 61, 62]; thus, the details are omitted here.

*3.4. Energy Consumption and Battery Depletion*

An energy consumption model updates the battery levels of EVs, including AVs and ETs, at each time step (0.1 s), based on their position relative to the CWD lane. Battery depletion is calculated at a rate of 0.29 kWh/mi ($1.8 \times 10^{-4}$ kWh/m) for AVs as reported by He et al. [38], with a battery capacity of 80 kWh. The battery of ETs is assumed to have a consumption rate of 2 kWh/mi ($1 \times 10^{-3}$ kWh/m) and a capacity of 1000 kWh [63]. In this study, the energy consumption is assumed to be proportional to the traveled distance. This assumption is valid for travel speeds under 100 km/hr [64] and is supported by multiple studies, including [65, 66].

Power gained on CWD lanes was modeled following Fuller [67], who suggested a power range of 0.33 to 3.33 kWh/min. We used the average value of 1.8 kWh/min for all AVs and ETs driving on CWD lanes. Drivers of EVs (including AVs and ETs) receive a high incentive in the MOBIL model to change lanes toward the CWD lane whenever their battery levels drop below 20% of battery capacity.

*3.5. Emission Modeling*

To assess the environmental impacts of CWD lanes we implemented an in-vehicle emission model based on Liu et al. [68], which calculates emission factors for four pollutants—Carbon Monoxide (CO), Nitrogen Oxides (NOx), Volatile Organic Compounds (VOCs), and Particulate Matter (PM2.5)—for different categories of vehicles, including medium passenger cars (MPCs) and heavy-duty trucks (HDTs), which we assumed correspond to the HVs and DTs considered in this study, at various speed levels using the COPERT IV model [69].

*3.6. Simulation Environment and Lane Configurations*

The simulation environment supports various lane configurations, including regular and CWD lanes, as well as on-ramps and off-ramps. CWD lanes require lower speed limits to function effectively and charge vehicles. To address this issue, Chen et al. [65] and Zhong et al. [66] considered an upper and lower limit for the speed of vehicles on CWD lanes. Zhong et al. [66] defined a range of (10 m/s to 20 m/s) for this matter, which is approximately $35 - 70$ km/hr. Deflorio et al. [70] assigned the CWD lanes to the right-hand side lane, as this lane has the slowest flow, with a maximum speed limit of 60 km/hr. Based on these studies, we assumed the lower initial speed range for all the vehicles generated into the simulation to be 60 km/hr, and the maximum speed limit is considered 120 km/hr, which is typical of interstate highways. The desired speed value for AVs and HVs is 120 km/hr and 80 km/hr for DTs and ETs.

---

[1]AV isolated sensor range.
[2]AV connected sensor range.
[3]System target time-gap.

## 4. Results

*4.1. Impact Analysis*

This section assesses the impacts of CWD lanes on traffic flow dynamics for different market penetration rates of EVs and AVs on a 5-km long three-lane highway segment with an on-ramp that is located 2.5 km from the start of the section. The traffic flow volume in the main lanes is $2,000$ vph/lane, and the on-ramp volume is 400 vph. This hypothetical three-lane highway with an on-ramp located in the middle of the segment is depicted in Figure2 with the CWD lane located at the rightmost lane and colored orange. The simulation duration of this segment is 5 minutes, with the first 150 seconds being the simulation's warmup period. The ramp speed is set to 60 km/hr, with a 300-meter-long acceleration lane.

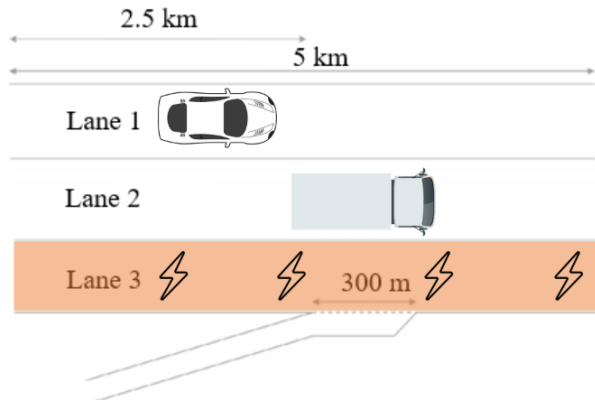

Figure 2: Highway segment under study.

Table 3 presents the different combinations of market penetration rates (MPRs) analyzed in this case study. We confined the EV's MPR to a maximum of 40%, which is around the predicted level to be reached by 2030 [1]. Although Miller et al. [71] predicted that the sales shares of medium- and heavy-duty zero-emission vehicles (ZEVs) will remain below 2% through 2050, we considered a maximum penetration rate of 10% for ETs to further investigate the impacts of CWD lanes under a more prominent presence of ETs, which benefit the most from these lanes. The first combination serves as a benchmark, while the last one represents the scenario where the impacts of CWD lanes are expected to be the highest.

An investigation of the fundamental diagram for different MPRs, depicted in Figure 3, indicates that higher MPRs of EVs (AVs + ETs) and DTs in the driving environment result in a decrease in the throughput rate and an increased scatter compared with the benchmark scenario. This is mainly due to the fact that DTs travel at lower speeds, and EVs tend to perform more lane-changing maneuvers to reach the CWD lane, which, in turn, causes more scatter and less throughput. However, at lower MPRs, where we introduce more AVs and fewer trucks, the throughput increases in the fundamental diagrams, which can be attributed to the enhancement caused by the more robust performance of AVs. Comparing AVs and ETs, we observe that introducing more ETs into the traffic flow results in a lower throughput rate, as evident when comparing MPRs of 30% and 40% because these vehicles tend to occupy the CWD lane and travel at a lower speed.

Table 3: Market Penetration Rates of the Vehicle Fleet

| Scenario | Market Penetration Rate | | | |
|---|---|---|---|---|
| | HV | AV | ET | DT |
| Base case | 100% | 0% | 0% | 0% |
| Case I | 90% | 5% | 0% | 5% |
| Case II | 80% | 10% | 5% | 5% |
| Case III | 70% | 20% | 5% | 5% |
| Case IV | 60% | 25% | 5% | 10% |
| Case V | 50% | 30% | 10% | 10% |

Figures 4 (a) and (b) suggest that when introducing DTs to the environment, the traffic flow shows significant shockwave formation; however, as the vehicles reach the on-ramp, they will proceed with higher speeds because the rightmost lane is not congested as the EVs MPR is still low and there is no incentivized lane changing behavior to reach the CWD lane at the right. As we move forward from Figure 4 (c) to (e), the MPR of AVs increases, and the shockwave patterns become more stable and localized.

Although, in Figure 4 (f), the dominance of AVs has improved speed harmonization and reduced disruptions and led to a more uniform traffic flow, the higher proportion of ETs continues to introduce localized disruptions in the CWD lane, as indicated by a reduction in speed around the on-ramp location. However, their impact is mitigated by the behavior of AVs.

The smoothed average speed per lane for different MPRs is presented in Figure 5. This figure shows a clear distinction in traffic flow dynamics when the CWD lanes are introduced, and EVs MPRs have increased; hence, it provides a good understanding of the impact of CWD lanes. In the benchmark scenario, Figure 5 (a), where EVs are absent, we can observe that the average speed in all of the lanes is nearly the same, and they have reached the highest value among all the scenarios (approximately 70 km/hr).

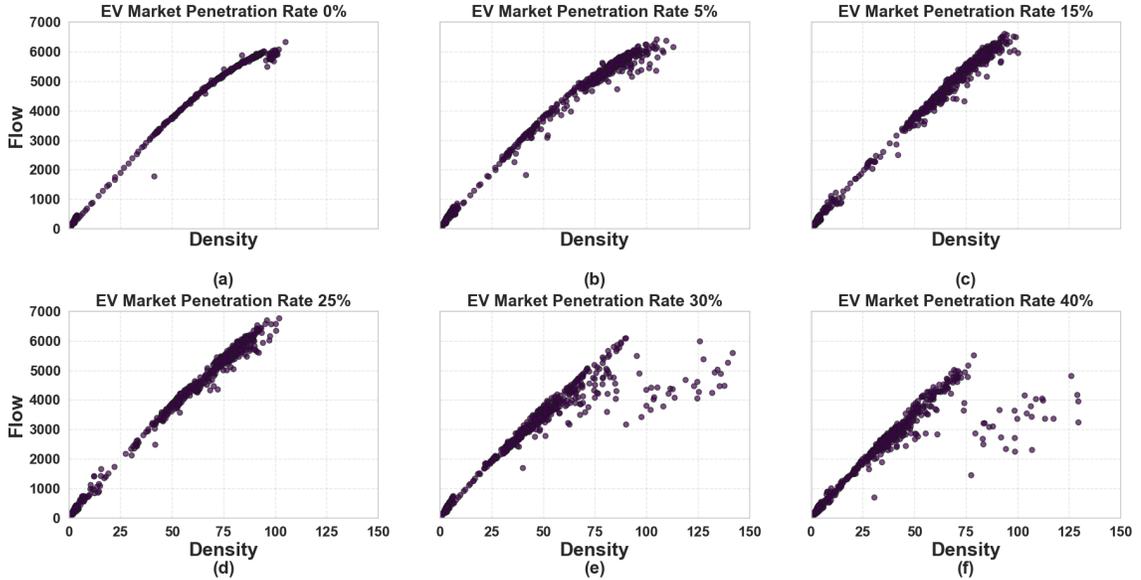

Figure 3: Fundamental Diagrams for Different MPRs: (a) Base Case: only HVs, (b) Case I: 5% AVs, (c) Case II: 10% AVs + 5% ETs, (d) Case III: 20% AVs + 5% ETs, (e) Case IV: 25% AVs + 5% ETs, and (f) Case V: 30% AVs + 10% ETs.

Furthermore, as the number of EVs increases, the variance between the average speed of different lanes becomes more distinct due to the growing presence of slower-moving ETs in the CWD lane (lane 3). Moreover, the disrupting impacts of slower-moving vehicles in the CWD lanes are not bound to the CWD lane. As we can observe, first, the average speed in lane 3 is dropped (as shown in figure 5 (d). Furthermore, the impacts of lane-changing maneuvers are propagated to lane 2 (adjacent to the CWD lane) and lane 1 as well, despite the better performance of AVs than HVs and the more efficient lane-changing process.

This phenomenon is most evident in Figure 5 (f), where the average speed over all lanes becomes unstable, with lane 3 having the lowest average speed over the most time steps. Compared with the initial values in Figure 5 (a),

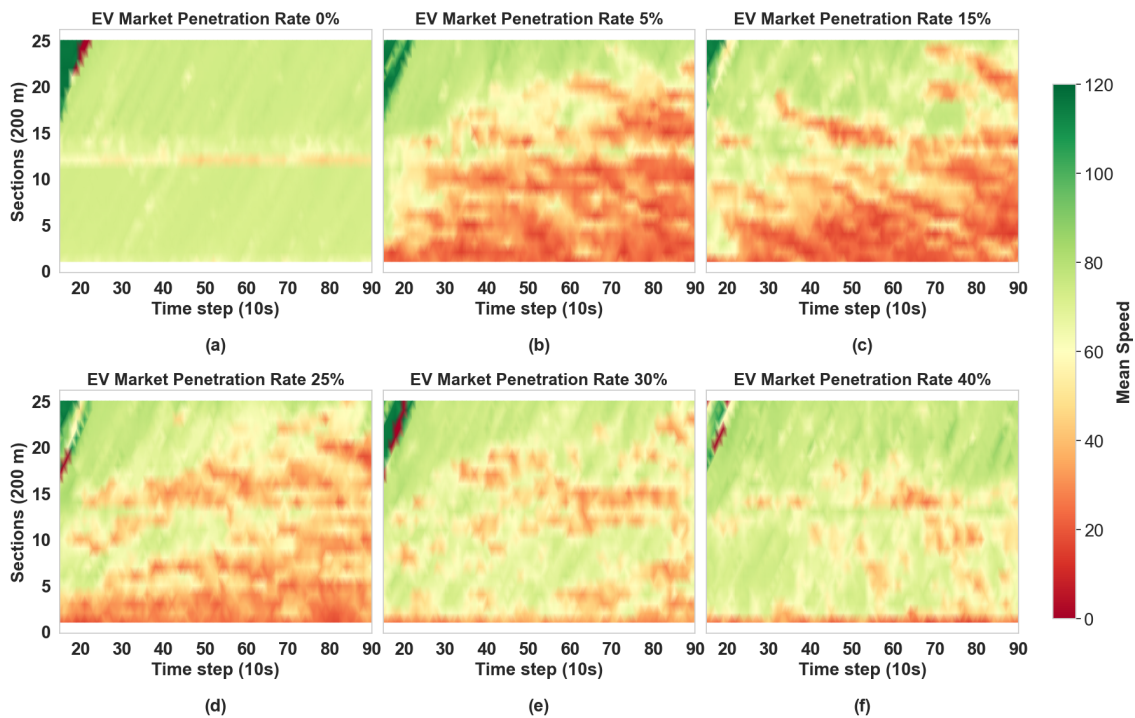

Figure 4: Traffic Shockwave Formation for Different MPRs in a 15-minute run: (a) Base Case: only HVs, (b) Case I: 5% AVs, (c) Case II: 10% AVs + 5% ETs, (d) Case III: 20% AVs + 5% ETs, (e) Case IV: 25% AVs + 5% ETs, and (f) Case V: 30% AVs + 10% ETs.

lane 3 maintains lower speed ranges (around the speed limit of the CWD lane).

Finally, the emission impacts of introducing EVs to the traffic flow are presented in Table 4, where the in-vehicle pollutant levels of CO, NOx, VOCs, and PM2.5 all decrease as EVs' MPR increases.

This brings valuable insight into the environmental benefits of EVs as indicated by Table 4, the CO emission has been reduced by 48.2%, the NOx Emissions have been reduced by 59%, VOC levels have been reduced by 63.3%, and PM2.5 levels have been reduced by 43.4%. Moreover, the rate of pollutant reduction is not uniform across MPRs. The steepest reductions are observed between 0% and 25% MPR, with diminishing marginal benefits beyond 25% MPR. Although EVs contribute to emissions reduction, their slower speeds in CWD lanes can slow down the entire traffic and can limit EVs' potential benefits (as the emission factors used in the COPERT IV model [69] are higher in lower speed ranges). This is evident from the slower reduction rates in PM2.5 and NOx emissions. Furthermore, the impact of introducing more ETs on emission rates is evident from diminishing reduc-tions from 25% to 40% MPRs for almost all pollutants in Figure 6, as these vehicles tend to occupy CWD lanes and travel at slower speeds (increasing emissions in other vehicle types), resulting in their increased MPRs having less effect on reducing emissions. Moreover, DTs have a significant impact on increasing emissions, as can be concluded by comparing the first and second columns in Table 4.

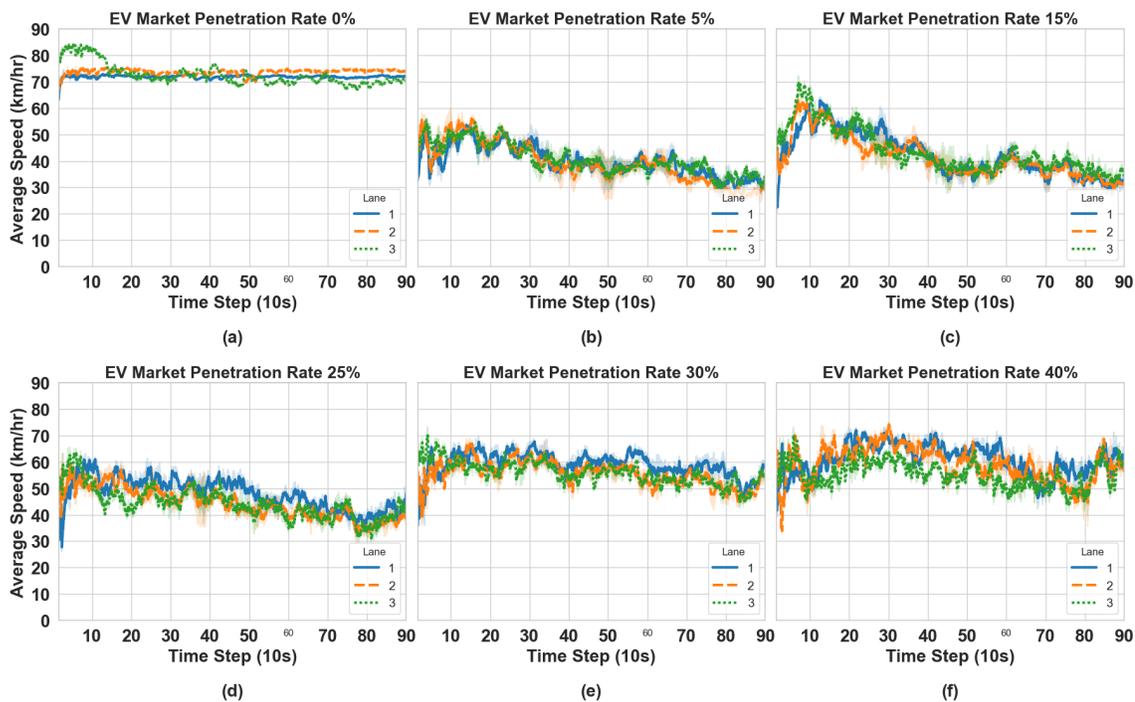

Figure 5: Smoothed Average Speed per Lane for Different MPRs while lane 3 is designated as the CWD lane in a 15-minute run: (a) Base Case: only HVs, (b) Case I: 5% AVs, (c) Case II: 10% AVs + 5% ETs, (d) Case III: 20% AVs + 5% ETs, (e) Case IV: 25% AVs + 5% ETs, and (f) Case V: 30% AVs + 10% ETs.

Table 4: Total Emissions (Tons) for Different EV MPRs based on the layout depicted in Figure 2

| Total Emissions | EV Market Penetration Rate | | | | | |
|---|---|---|---|---|---|---|
| | **0%** | **5%** | **15%** | **25%** | **30%** | **40%** |
| **CO** | 1924.77 | 1750.18 | 1355.73 | 1250.13 | 1183.22 | 996.24 |
| **NOx** | 1389.84 | 1268.00 | 1075.79 | 964.05 | 690.99 | 569.92 |
| **VOCs** | 1205.22 | 1052.79 | 854.67 | 760.10 | 582.24 | 442.63 |
| **PM2.5** | 15.20 | 14.86 | 12.13 | 11.29 | 9.20 | 8.61 |

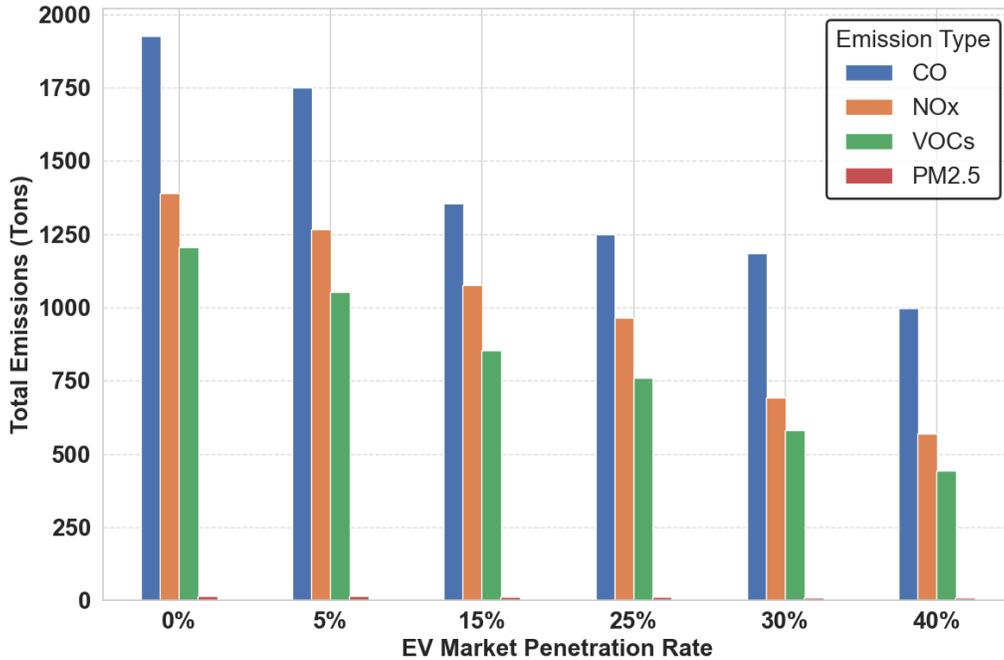

Figure 6: Emission levels for different EV MPRs based on the layout depicted in Figure 2.

## 4.2. Design Policy

This section is dedicated to analyzing different design policies for implementing CWD lanes on a highway. To achieve this objective, we considered different designs, configurations, and slopes of CWD lanes and compared the results for each MPR to give roadway designers and policymakers better insight. The details are mentioned in Table 5. It is important to note that the presented design alternatives are not a comprehensive list of possible scenarios. The presented alternatives are selected based on existing design approaches in real-world CWD lanes experiments.

Table 5: Policies for CWD Lane Configurations

| Policy | Description |
|---|---|
| F1 | Full-length CWD lane implemented throughout all sections in lane 1. |
| F2 | Full-length CWD lane implemented throughout all sections in lane 2. |
| F3 | Full-length CWD lane implemented throughout all sections in lane 3. |
| C1.O | CWD lane in lane 1 cut at the on-ramp (section 13). |
| C1.A | CWD lane in lane 1 cut at and after the on-ramp (sections 13 and 14). |
| C1.B | CWD lane in lane 1 cut at and before the on-ramp (sections 12 and 13). |
| C2.O | CWD lane in lane 2 cut at the on-ramp (section 13). |
| C2.A | CWD lane in lane 2 cut at and after the on-ramp (sections 13 and 14). |
| C2.B | CWD lane in lane 2 cut at and before the on-ramp (sections 12 and 13). |
| C3.O | CWD lane in lane 3 cut at the on-ramp (section 13). |
| C3.A | CWD lane in lane 3 cut at and after the on-ramp (sections 13 and 14). |
| C3.B | CWD lane in lane 3 cut at and before the on-ramp (sections 12 and 13). |
| S1 | Full-length CWD lane implemented throughout all sections in lane 1 on a +3% sloped terrain. |
| S2 | Full-length CWD lane implemented throughout all sections in lane 2 on a +3% sloped terrain. |
| S3 | Full-length CWD lane implemented throughout all sections in lane 3 on a +3% sloped terrain. |

The "F" policies study the impacts of CWD lanes when installed on all sections of the highway regardless of the merging effects of entering flow from the on-ramp. Note that the F3 policy was investigated in the previous section. The policy number indicates the lane where CWD infrastructure is placed. The "C" policies consider the scenarios where the CWD lane is truncated adjacent to the on-ramp to see if any performance improvement can be achieved. To further investigate this, we defined three sub-policies where the CWD lane is truncated precisely at the section matching the entry point of the on-ramp ("O"), a section (200 m) after the end of the merging area ("A"), or a section (200 m) before the start of the on-ramp ("B"). The "S" policies aim to study the impacts of a sloped terrain on a system equipped with a full-length CWD lane. The "F" and "C" policies are studied first, as they both reflect a flat highway. A sample layout for different policies applied to lane 3 in our hypothetical highway is shown in Figure 7. We considered different layouts with respect to the merging point, as this is expected to be one of the key sources of disturbance in the main lanes. Note that such a setup allows EVs to change to other lanes (to avoid merging vehicles) without losing any charging opportunities in the vicinity of the on-ramp.

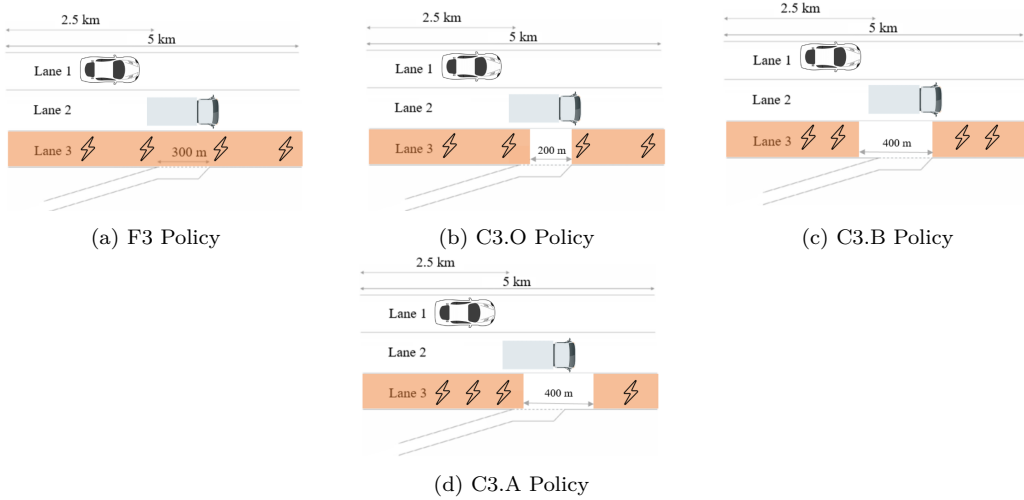

Figure 7: CWD Lane Design in Lane 3; identical configurations apply to other lanes with the CWD lane shifted accordingly.

Figure 8 presents the average speed of vehicles (throughout the simulation) in all lanes for different policies across different MPRs. The results indicate that:

- First, the full-length policies (F) perform worse than those in which the CWD lane is truncated near the on-ramp.

- Second, in almost all of the MPRs, installing the Connected and Wireless Driving (CWD) lane in lane 3 would enhance the performance. as

illustrated in Figure 8 (a), when the MPR of electric vehicles (EVs) is low, the average speed of vehicles in lane 2 is relatively higher than that in lane 3. In this scenario, the majority of the fleet consists of HVs. Unlike AVs, HVs do not receive information about merging vehicles from the on-ramp through the infrastructure. Consequently, they may be affected by vehicles merging from the on-ramp, especially if lane 3 is congested. Therefore, it is advisable to move the CWD lane to lane 2 and truncate it at the merging section (Policy C2.O). In summary, it is generally best to install the CWD lane in lane 3, except in instances where the MPR of EVs is low.

- Third, as the MPR of EVs increases, the performance of the whole system improves, and the average speed across all of the policies increases.

In Figure 9, the total energy supplied by the CWD lanes is assessed for different policies and MPRs. This figure shows that full-length policies (F) provide more chances for EVs to charge their batteries; hence, the energy supply in these scenarios is higher than other policies. This further demonstrates the paradox between the benefits provided by CWD lanes to EVs and their impact on traffic flow dynamics, as average speeds in F policies are low, but the total energy supplied is highest in these policies. Moreover, Figure 9 (e) demonstrates that as ETs MPR increase, less charging opportunity is provided for vehicles as ETs tend to occupy CWD lanes; therefore, the energy supplied to EVs is reduced in this scenario.

We compared the total and normalized emission levels for different policies in Figures 10 and 11, respectively. The results demonstrate that the NOx, VOCs, and PM2.5 pollution levels are more or less independent of the policy and CWD lane design. However, CO emission tends to rise in F policies.

Eventually, the effect of a 3% up-ward slope on the mixed traffic flow in the presence of a full-length CWD lane is assessed by considering the weight and friction resistance of DTs and ETs. We ignored these effects on HVs and AVs as these are passenger vehicles. The acceleration rates are adjusted as follows in this scenario:

$$
\begin{aligned}
a_\text{adjusted} &= a_\text{IDM} - g \cdot \sin(\theta) - \mu \cdot \cos(\theta) \\
&\simeq a_\text{IDM} - g \cdot \left(\theta + \mu \cdot \left(1 - \frac{\theta^2}{2}\right)\right)
\end{aligned}
\tag{6}
$$

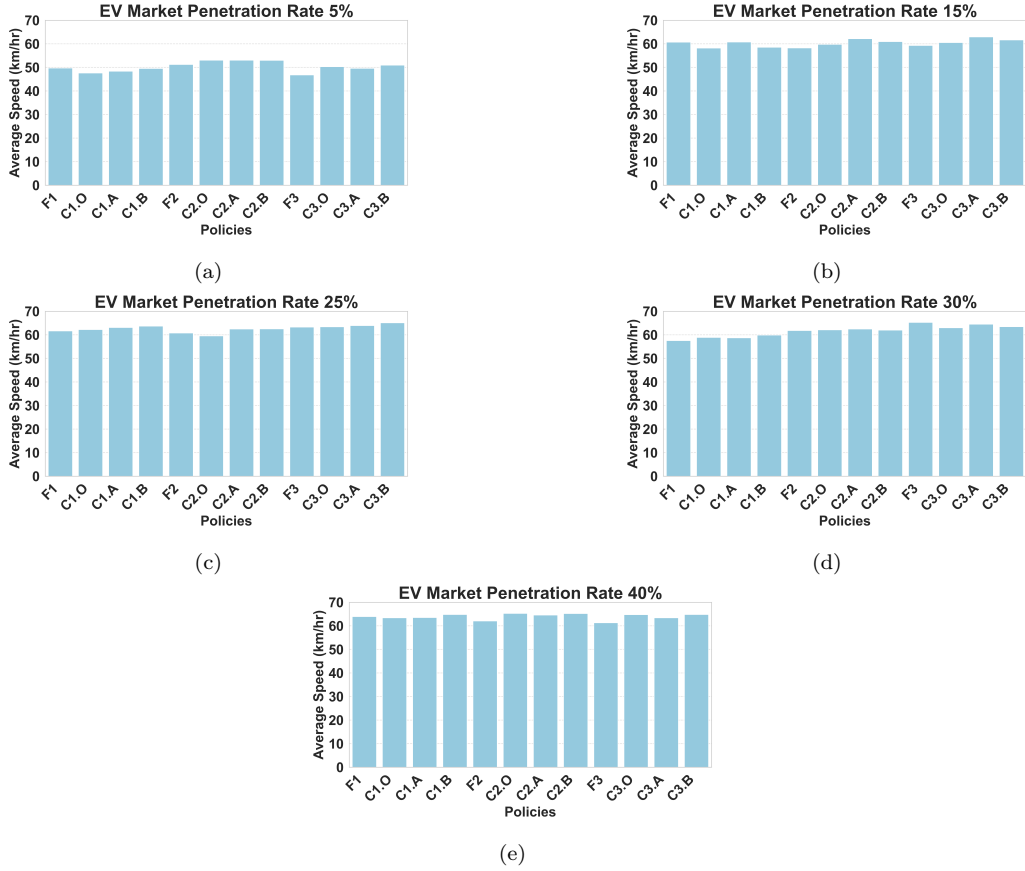

Figure 8: Average Speeds for Different Policies Across Various MPRs: (a) Case I: 5% AVs, (b) Case II: 10% AVs + 5% ETs, (c) Case III: 20% AVs + 5% ETs, (d) Case IV: 25% AVs + 5% ETs, and (e) Case V: 30% AVs + 10% ETs.

where $a_{\text{IDM}}$ is the acceleration rate derived from the IDM model, $g$ is the acceleration due to gravity (9.8 m/s$^2$), and $\mu$ is the friction factor which is equal to 0.8 for a dry asphalt road [72].

The average speed depicted in Figure 12 indicates that in a sloped terrain, the system's performance deteriorates due to sluggish trucks. Moreover, DTs have a more significant impact as these vehicles, which already had lower acceleration rates, now have to struggle even more to keep up with the traffic flow. This is evident from the performance deterioration compared with flat terrain scenarios.

The total electricity consumed by EVs in the sloped scenario is depicted in Figure 13. This figure shows that the energy consumption has increased in the sloped scenario because ETs face reduced acceleration and speed. Moreover, if increasing battery levels is the top priority, in almost all MPRs, it is best to install CWD lanes on the leftmost lane (lane 1) as EVs can have more charging opportunities.

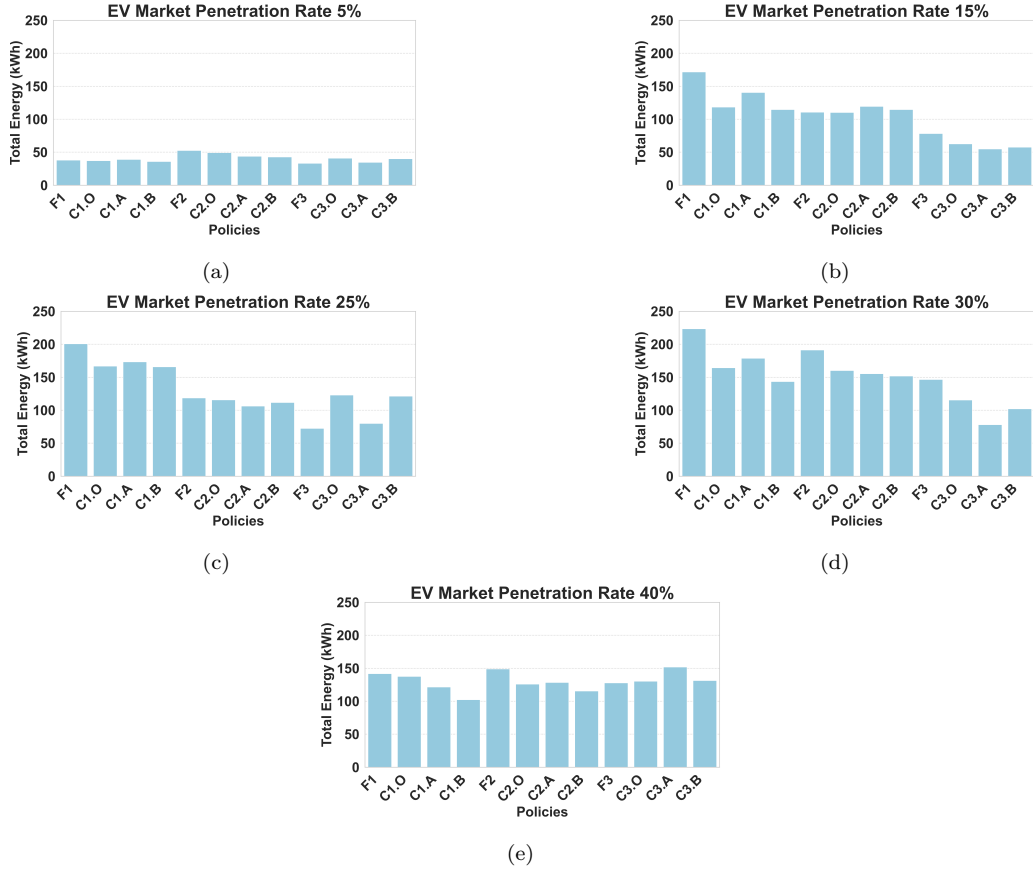

Figure 9: Total Energy Consumption for Different Policies Across Various MPRs Over 5 Minutes of Simulation: (a) Case I: 5% AVs, (b) Case II: 10% AVs + 5% ETs, (c) Case III: 20% AVs + 5% ETs, (d) Case IV: 25% AVs + 5% ETs, and (e) Case V: 30% AVs + 10% ETs.

## 5. Discussion

Evaluating the results from the fundamental diagram, shockwave formation patterns, and average speed for each lane indicates the impacts of CWD lanes on overall traffic flow patterns. The observed reduction in overall traffic throughput and increased scatter in the fundamental diagram at higher EV MPRs underscores the disruptive nature of frequent lane-changing maneuvers by EVs aiming to access the CWD lane. While AV technologies exhibit the potential to reduce these disruptions, their effectiveness diminishes with higher ET presence, highlighting the importance of optimized lane-management strategies and advanced AV algorithms in mixed-vehicle environments. Moreover, the beneficial impact of higher AV MPRs on mitigating shockwave formations indicates the critical role of automation in stabilizing traffic flow. Nevertheless, the lane-changing behavior of EVs aiming to reach the CWD lane disrupts on-ramp merging efficiency, revealing challenges in coordinating between infrastructure-based charging and traffic operational goals.

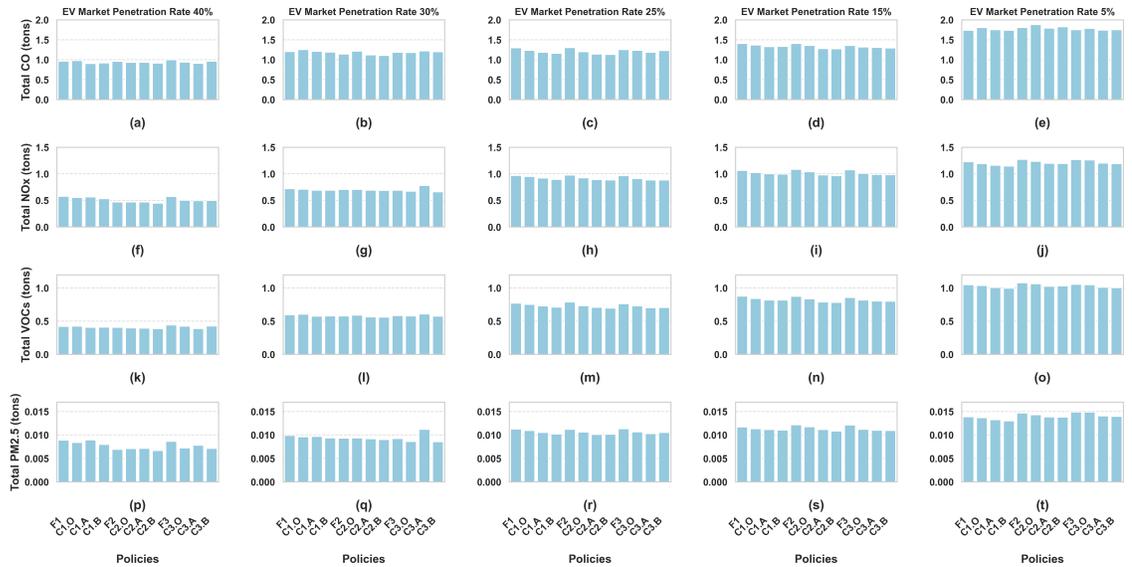

Figure 10: Emission Levels for Different Policies Across Various MPRs: (a, f, k, and p) Case I: 5% AVs, (b, g, l, and q) Case II: 10% AVs + 5% ETs, (c, h, m, and r) Case III: 20% AVs + 5% ETs, (d, i, n, and s) Case IV: 25% AVs + 5% ETs, and (e, j, o, and t) Case V: 30% AVs + 10% ETs.

From the average speed perspective, the substantial reduction in average speeds within the CWD lane and adjacent lanes at higher EV MPRs illustrates the unintended consequences of policy-induced lane preferences. Particularly, ETs contribute significantly to the congestion due to their size and lower speeds. This emphasizes the need for targeted infrastructure design and vehicle-type management to prevent the exacerbation of traffic issues. Finally, the considerable reductions in CO, NOx, VOCs, and PM2.5 emissions at lower EV MPRs confirm the environmental benefits of EV promotion policies. However, diminishing returns observed beyond 25% MPR indicate that congestion associated with high EV MPRs can partially offset expected environmental gains. Thus, effective congestion management strategies are essential to maintain these environmental benefits at higher EV adoption levels.

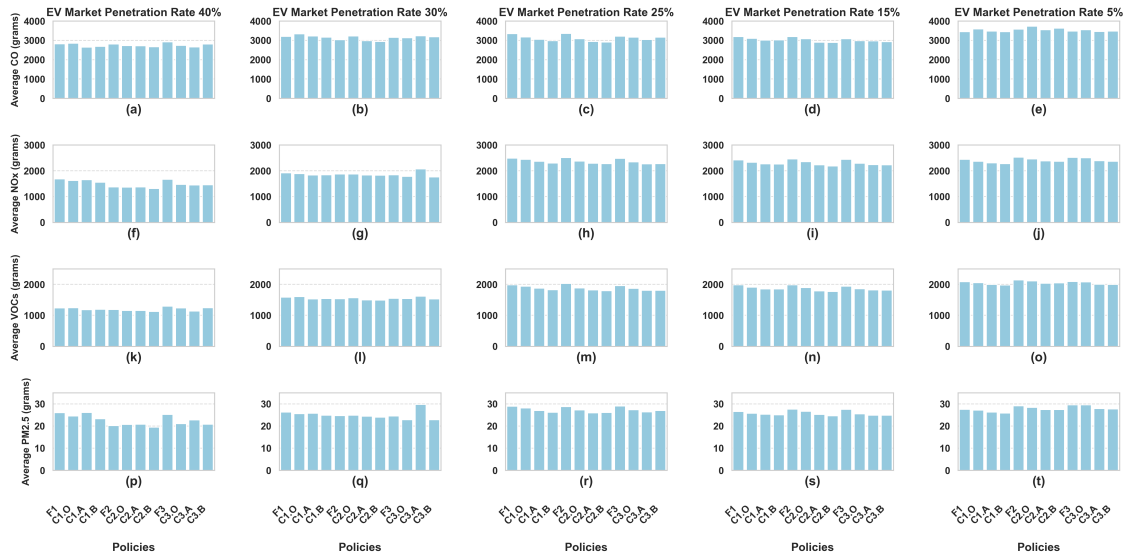

Figure 11: Normalized Emission Levels for Different Policies Across Various MPRs: (a, f, k, and p) Case I: 5% AVs, (b, g, l, and q) Case II: 10% AVs + 5% ETs, (c, h, m, and r) Case III: 20% AVs + 5% ETs, (d, i, n, and s) Case IV: 25% AVs + 5% ETs, and (e, j, o, and t) Case V: 30% AVs + 10% ETs.

The analysis of fifteen distinct CWD lane policies highlights an inherent trade-off: while CWD lanes effectively address range anxiety issues for EV users, they simultaneously create disruptions in traffic flow. Our findings suggest that as AV penetration increases, strategically placing CWD lanes on the rightmost lane and limiting their length near on-ramp merging points can optimize outcomes. Conversely, full-length installation is preferable if the primary policy objective is reducing dependency on charging stations. Additionally, emissions analysis reveals that CO levels are sensitive to CWD lane length, whereas NOx, VOCs, and PM2.5 are relatively unaffected, em-phasizing the nuanced impact of infrastructure design on specific pollutants. Finally, installing CWD lanes on sloped terrains requires careful consideration of policy goals, as rightmost lanes maximize traffic speeds, while leftmost lanes enhance EV charging opportunities.

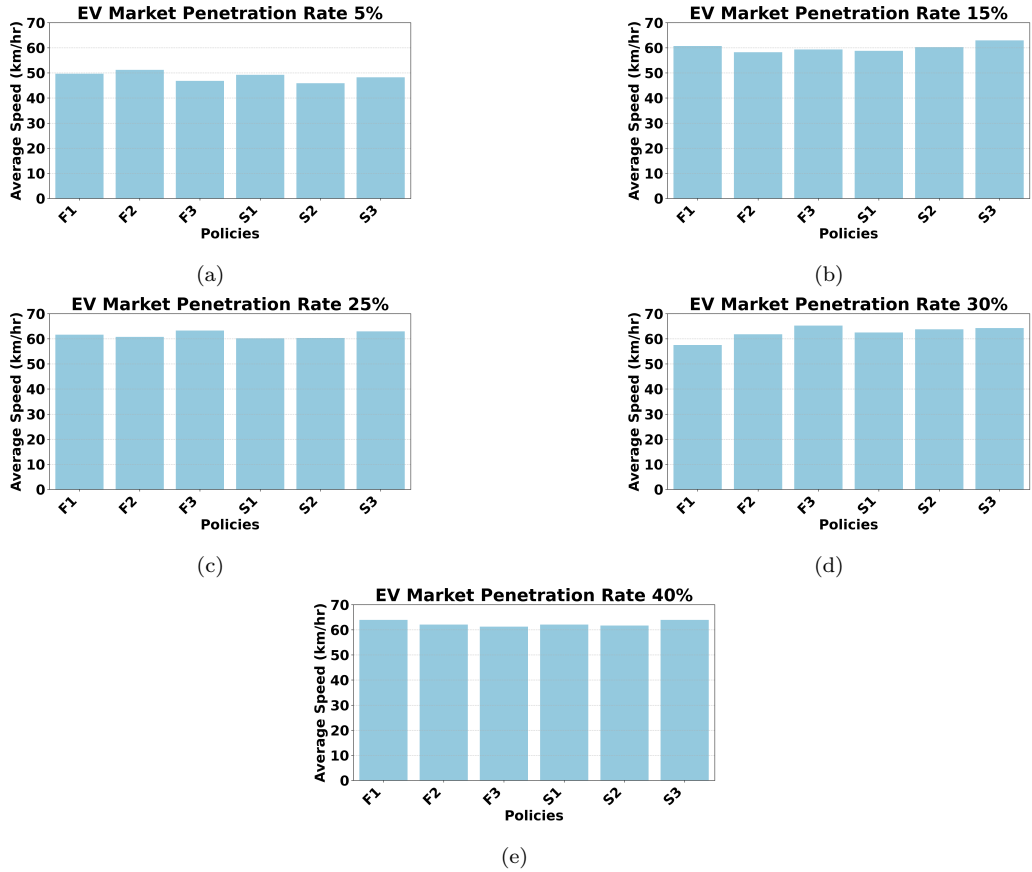

Figure 12: Average Speeds for Different Policies Across Various MPRs: (a) Case I: 5% AVs, (b) Case II: 10% AVs + 5% ETs, (c) Case III: 20% AVs + 5% ETs, (d) Case IV: 25% AVs + 5% ETs, and (e) Case V: 30% AVs + 10% ETs.

## 6. Conclusion

This study explored the implications of integrating CWD lanes within mixed traffic environments, emphasizing the balance between supporting EV adoption and maintaining efficient traffic operations. Our analyses underscored both the potential and limitations of current CWD lane designs. From a congestion perspective, higher EV MPRs resulted in more scatter in the fundamental diagram that could only be offset to some degree at higher penetration rates of AVs. Moreover, the simulation results indicated that as the MPR of EVs increases, emissions do not necessarily decrease as more congestion is expected at higher penetration rates of EVs. Finally, various CWD design policies were evaluated, and it was found that there is a need for coordination between infrastructure-based charging and traffic operational goals.

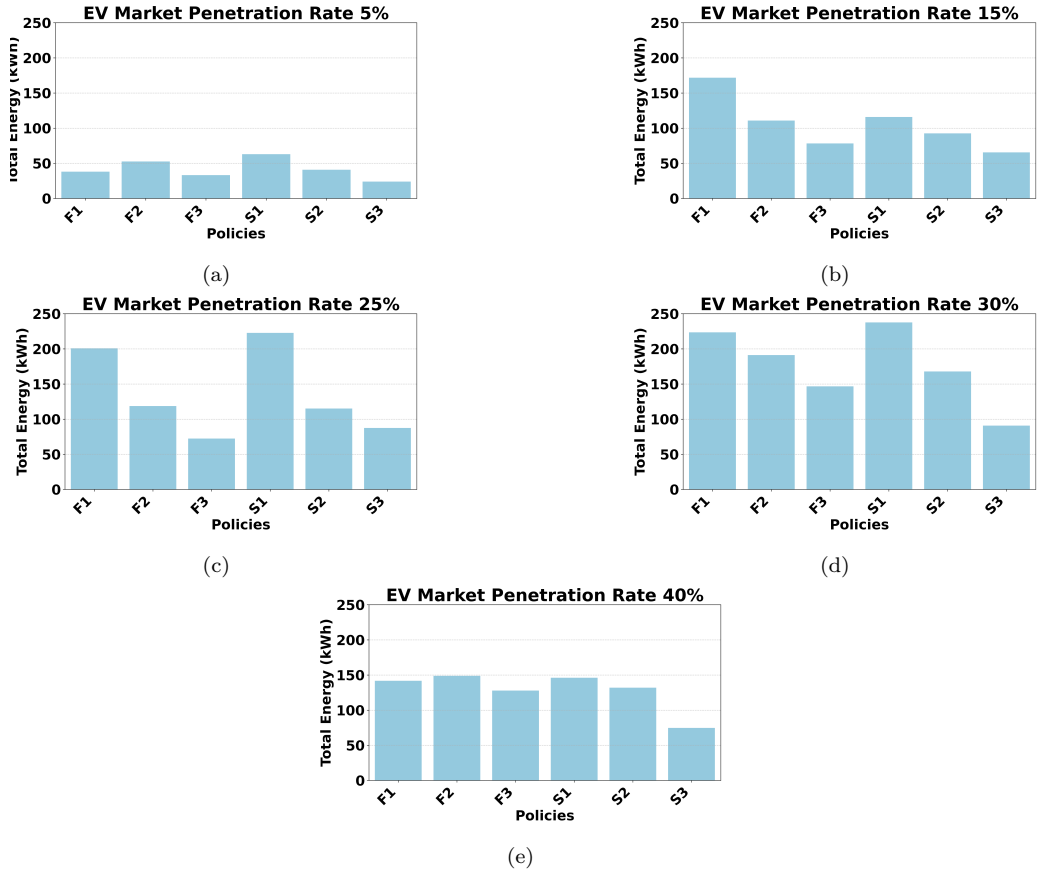

Figure 13: Total Energy Consumption for Different Policies Across Various MPRs Over 5 Minutes of Simulation: (a) Case I: 5% AVs, (b) Case II: 10% AVs + 5% ETs, (c) Case III: 20% AVs + 5% ETs, (d) Case IV: 25% AVs + 5% ETs, and (e) Case V: 30% AVs + 10% ETs.

Implementing CWD lanes involves significant opportunities and inher-ent challenges. To achieve optimal outcomes, policymakers should carefully align infrastructure designs with specific goals regarding traffic efficiency, environmental benefits, and energy management. Future research should focus on optimizing the design of CWD lanes by developing and analyzing more policies, emphasizing the need for strategic planning in allocating CWD lanes and managing the mix of vehicle types that utilize them. Performing a comprehensive cost-benefit analysis of installing CWD lanes can provide policymakers with better insights into constructing new charging stations or CWD lanes.

**Ethical Statement**

The authors declare that they have no known competing financial interests or personal relationships that could have appeared to influence the work reported in this research.